\documentstyle[11pt,newpasp,twoside]{article}
\def\edcomment#1{\iffalse\marginpar{\raggedright\sl#1\/}\else\relax\fi}
\reversemarginpar

\begin{document}
\title{Prospects of Variable Star Research by Future Space Missions}
 \author{Laurent Eyer}
\affil{Instituut voor Sterrenkunde,
             Katholieke Universiteit Leuven,
             Celestijnenlaan 200 B, B-3001 Leuven,
             Belgi\"e}

\begin{abstract}
ESA and NASA are studying projects having a tremendous return on variable
star research. Other national space agencies are also studying or developing
projects of smaller costs but with impressive returns.
The projects range from global Galactic surveys like the ESA mission GAIA
which will give photometric time series for about 1 billion
stars, to detailed pulsation modes studies like the CNES
mission COROT which could reach a photometric precision lower
than 1 ppm. The presentation will emphasize
the future astrometric, asteroseismologic and planet
detection missions.
\end{abstract}

\section{Introduction}
Many advantages are present when going in space. The
most exploited one has been the access to wavelengths not
transmitted through the atmosphere.
For photometry near the visible part of the spectrum, there are
many benefiting aspects:
No atmosphere means no absorption bands of $H_2O$ and $O_3$,
no scintillation, no emission or diffusion from the sky (dark
sky), no variable extinction.
Furthermore depending on the satellite orbit, it may be
possible to make long continuous monitorings (however eclipses
by the Moon or the Earth may occur) or  to
make a global sky coverage.

As a comparative example, the COROT team (see section~6)
studied fictive Earth projects (cf. Baglin 1997a) and showed
that even an immoderate project of 8 8m-class telescopes
disseminated around the Earth would not surpass their small
space mission.

The drawbacks of space projects are their complexity, cost,
aging, limited operation time, and many technical constraints.
Space missions are fragile on technical and also political
point of views. The global budgets of ESA and NASA in local
currencies have been relatively flat for about 8 years, but
they may be subjected to high volatility according to changes
in the policy.
\section{Main types of missions}
Three main types of missions will be presented: one
performing astrometry, one oriented towards asteroseismology
and one aiming at detecting telluric planets. Many of these projects
lead to massive photometric surveys and have thus very valuable
scientific returns as discussed by Paczy\'nski (1997).
Then, an overview of the distinctive features of
some specific satellites will be given.
%
\subsection{Astrometric missions}
After the pioneering mission Hipparcos, projects performing
astrometric measurements are flourishing. Thanks to the advance
of technology (CCD, interferometry), the jump in
astrometric precision is of 2-3 orders of magnitude, thus making
a new mission very attractive. The accuracies reached by these
projects are now labeled in $\mu$as (micro arcsec), cf. Table~1.

The interest for variable stars is threefold.
First, several proposed missions are scanning the whole sky and
performing photometry. They will thus give, as Hipparcos did,
huge amounts of photometric measurements, achieving magnitude
completeness. As the scanning law
is optimised for the astrometric purpose, the time series
have a semi-regular sampling which produce aliasing.
Second, the information of parallax is a stringent constraint 
for testing stellar models when used in asteroseismology (cf.
Baglin 1997b, Favata 1999).
In the case of Hipparcos for instance, Matthews et al. (1999)
showed a comparison of asteroseismologic parallax versus
trigonometric parallax. 
On an other hand, absolute luminosities or masses derived from
parallaxes can be used as starting point of the seismologic
models.
Third, Parallaxes are also of primordial importance for calibrating
distance indicators as Cepheids, RR Lyrae and Miras. All the
presented missions will literally pin down many of these
objects.
\begin{table}[htb]
 \caption{Summary of several experiments performing astrometry.
 Mag is the magnitude
 range or limiting magnitude, N the number of measured stars in
 million, $\sigma_A$ the astrometric precision in $\mu$as at
 $V=9$ and $V=15$, $T_D$ is the date for realisation decision
 (Month/Year), $T_L$ is the year of the launch.}
 \begin{center}
   \begin{tabular}{|lrrrrrr|} \hline 
Mission& Agency     & Mag &    N& $\sigma_A$ & $T_D$ & $T_L$ \\ \hline
   GAIA& ESA        &   20& 1000&   4,  10   & 3/00& 2009    \\
   FAME& USNO/NASA  & 5-15&   40&  50, 500   & -/99& 2003/2004\\
   DIVA& Germany    &   15&   35& 150, 5000  &12/99& 2003     \\ \hline
   \end{tabular}
 \end{center}
\end{table}
\subsection{Asteroseismologic missions}
Several unsuccessful attempts have been made to measure solar-like
oscillations from Earth, for such measurements the precision in
luminosity determination to be reached is as high as few ppm,
furthermore high frequency resolution is required which imposes
very long continuous monitoring.
As  previously remarked, the scintillation is a limiting factor
which is magnitude independent as long as photon noise is not
dominating. With space missions these requirements can be reached.
\subsection{Planet detection missions}
Among several methods to detect planets, the most famous is the
measurement of the periodic variation in radial velocity that
a planet induces on its host star. Another promising technique,
shown to be more appropriate for the discovery of telluric
planets, is the detection of the luminosity decrease during an
eclipsing transit of the planet.
In this case, the system planet-star should be in a favorable
configuration for the observer. Thus large numbers of stars
must be observed to achieve valuable statistics.
\section{The GAIA Mission}
The goal of the GAIA
(\verb==http://astro.estec.esa.nl/SA-general/Projects/GAIA/)
satellite is to describe our Galaxy and
its history by making an inventory of 1\% of its stars
(down to magnitude V=20) in terms of their kinematical and
physical properties. However, the impact of GAIA will go well
beyond gaining knowledge of the Galaxy, it will infiltrate many
domains of astronomy such as stellar astronomy, general
relativity and cosmology. GAIA should not be seen merely as a
Super-Hipparcos; it has several instruments among them one is
performing astrometry reaching  the precision of 4 $\mu$as for
magnitude $V=12$ and  10 $\mu$as for magnitude $V=15$. These
measurements lead to parallaxes and proper motions.
On board, GAIA has a spectrometer which measures radial velocities
(with precision of 3-10 km/s for $V < 17$) to complete the 3-dim
velocity field.  A provisional 11 intermediate band photometric
system was specially designed in order to determine the
temperature, gravity and metallicity of
the observed stars. GAIA will produce photometric time series of about
100 measurements spread over 5 years.  The precision is magnitude
dependent (cf. Grenon et al. 1999) and for a single transit is of
the order of 0.05 mag at $V=17$ in the band F51 (centered at
570 nm and of 90 nm width).
The time sampling is semi-regular as a result of the scanning law
of the satellite.
Eyer \& Cuypers (these proceedings) are studying the expectation for
variable star numbers. Although the estimations are at preliminary
stages, GAIA might detect  a few thousand Cepheids and about
90\,000 RR Lyrae. GAIA will permit to study star pulsations
in terms of the physical parameters, especially the metallicity.
\section{FAME}
FAME (\verb==http://aa.usno.navy.mil/FAME/) stands for
Full-sky Astrometric Mapping Explorer. FAME will observe 40
million stars and will perform a Galactic survey magnitude
complete up to $V=15$. The astrometric precision will be $<$ 50
$\mu$as for mag $V<9$ and $<$ 500 $\mu$as for mag $V<15$.
It will collect photometric measurements in four bands of the
Sloan system (Fukugita et al. 1994). The announced precision of the
photometry is 1.6~mmag for mag 9 and 25~mmag for mag 15 in the
astrometric filter.
The scanning law will determine the time sampling. The rotation
period of the satellite is 40 minutes, the length of the
mission is 2 years and a half. The satellite will furnish
thousands of measurement per star, with a semi-regular
sampling grouped in sequences of 9-31-9 etc minutes.
\section{DIVA}
The DIVA (\verb=http://www.aip.de/groups/DIVA/=, Double
Interferometer for Visual Astrometry) satellite is a project of
the German space agency (DLR). 
DIVA will reach an astrometric precision of 150 $\mu$as at $V=9$
and 5 mas at $V=15$.
DIVA will have broad-band and
narrow-band systems. The photometric precision is under study.
The scanning law is very similar to the Hipparcos one, with a
rotation period of 2 hours.
\vspace{-0.5truemm}
\section{The COROT Mission}
The COROT (\verb=http://www.astrsp-mrs.fr/www/pagecorot.html=,
COnvection and ROtation) satellite has two major goals, first
it was designed for doing asteroseismology, focussing on 5 principal
fields for very  precise and continuous monitoring of 5 months each.
The noise level is 0.6 ppm for G type stars over 5 days.
The long time base makes possible a high accuracy on frequencies
determination (0.1 $\mu$Hz) necessary for the detailed seismic
investigation of the stellar structure proposed for COROT.
A planet detection program using the transit method was added,
this other goal has to survey large sample of stars (30~000 to
60~000) with a photometric precision of 0.1~\% for 16 minutes
of integration time in order to
detect telluric planets. This survey is achieved in dense
fields down to mag 15. Two-color information will be available
for $\sim$ 1/3 of the targets to establish the achromaticity
of the phenomenon.
There is also an exploratory program where COROT will measure
50 to 80 solar-like stars with $V < 9$ at a precision better than
2.4~ppm.
\vspace{-0.5truemm}
\section{MONS}
MONS (\verb=http://www.obs.aau.dk/MONS/=, Measuring
Oscillation in Nearby Stars) is a project of the Danish Small
Satellite Program, which already launched one successful
satellite ({\O}rsted).
MONS will observe about 20 solar-like stars over two years as a
primary scientific objective; it will also measure $\delta$
Scuti, roAp stars and samples of stars of all types as a
secondary objective. Its
eccentric orbit with high apogee (Molniya orbit) permits
access to a large part of the sky. It will measure changes in
colour ratio. Oscillations of amplitude of 1-10 ppm
will be detected (cf. Kjeldsen et al. 1999).
Furthermore, MONS will obtain science data
from the Star Imagers designed for attitude
control; they will permit measurement of about 900 stars,
700 in with the Imager SI 1, and 200 with the Imager SI 2.
The accuracy will be about 11 ppm (SI 1) and 26 ppm (SI 2)
at magnitude $V=6$.
\vspace{-0.5truemm}
\section{MOST}
MOST (\verb=http://www.astro.ubc.ca/MOST/=, Microvariability
and Oscillations of STars) is a funded
project which is the first satellite of the
Canadian Space Agency. MOST will observe several (3-5) solar-like
stars over 40 days. It will also observe roAp,
Wolf-Rayet and $\delta$ Scuti stars.
The photometric precision for a $V=6$ star is better than a few
ppm for 10 days of integration.
\section{Kepler}
Although the Kepler project (\verb=http://www.kepler.arc.nasa.gov/=)
was not selected due to questions about the ability to perform
the photometry in space,  the scientific case was
rated very highly. The Kepler team is planning to
resubmit the project.
Its goal is to "explore the structure and diversity of planetary
systems" by performing differential photometry.
It will monitor 100\,000 main-sequence stars in the
magnitude range from 9th to 14th with an accuracy of 0.002~\%
including instrument noise, shot noise and stellar variability
for a star of $V=12$. The mission would point continuously at a
single field in Cygnus during 4 years.
Because only data for preselected stars are saved,
all objects to be monitor must be pre-specified.
The team will probably entertain "guest observing" for
additional interesting objects like variable stars.
A subset of a few hundred brighter stars will also be
incorporated and observed with a shorter integration time
for asteroseismological purposes.

\section{Conclusion}
Many of these projects might be selected in a near future for
their realisation.
The projects performing astrometry will provide deep galactic
surveys and thus an inventory for galactic variable stars.
The asteroseismological projects have common goals but they
use different ingenious techniques and have different and diverse
by-products.

The future spatial projects will open new domains in variable
star studies, therefore precise predictions are delicate.
Since some presented missions are already funded, there
is the conviction that after the projects like Hipparcos
or the microlensing surveys, the domain of variable stars will
still be under a strong evolution.

\end{document}